\documentclass[preprints,communication,accept,pdftex, oneauthor]{Definitions/mdpi} 

\firstpage{1} 
\pubvolume{1}
\issuenum{1}
\articlenumber{0}
\pubyear{2023}
\copyrightyear{2023}
\datereceived{} 
\daterevised{ } 
\dateaccepted{ } 
\datepublished{ } 
\hreflink{https://doi.org/10.3390/universe9070298} 
\pdfoutput=1 

\Title{Femtoscopic correlation measurement with symmetric L\'evy-type source at NA61/SHINE}

\TitleCitation{Title}

\Author{Barnab\'as P\'orfy$^{1,2}$\orcidA{} on behalf of the NA61/SHINE Collaboration }

\AuthorNames{Firstname Lastname, Firstname Lastname and Firstname Lastname}

\AuthorCitation{Lastname, F.; Lastname, F.; Lastname, F.}

\address{%
$^{1}$ \quad Wigner Research Centre for Physics, Konkoly-Thege Mikl\'os \'ut 29-33, H-1121 Budapest, Hungary;porfy.barnabas@wigner.hu \\
$^{2}$ \quad Department of Atomic Physics, Faculty of Science, E\"otv\"os Lor\'and University, P\'azm\'any P\'eter s\'et\'any 1/A, H-1111 Budapest, Hungary; barnabas.porfy@cern.ch}

\abstract{Measuring quantum-statistical, femtoscopic (including final state interactions) momentum correlations with final state interactions in high-energy nucleus-nucleus collisions reveal the space-time structure of the particle-emitting source created. In this paper, we report NA61/SHINE measurements of femtoscopic correlations of identified pion pairs and describe said correlations based on symmetric L\'evy-type sources in Ar+Sc collisions at 150\textit{A} GeV/\textit{c}. We investigate the transverse mass dependence of the L\'evy-type source parameters and discuss their possible interpretations.}


\begin{document}

\section{Introduction}
\begin{figure}
\includegraphics[width=1.0\linewidth]{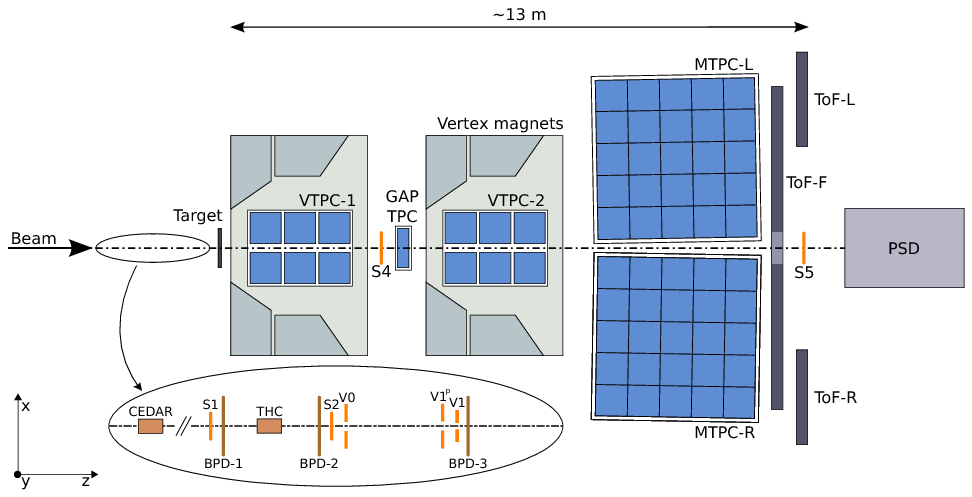}
\caption{The setup of the NA61/SHINE detector system during the run of Ar+Sc.}
\label{f:na61etup}
\end{figure}

The NA61/SHINE is a fixed target experiment using a large acceptance hadron spectrometer located in the North Area H2 beam line of the CERN Super Proton Synchrotron accelerator~\cite{Abgrall:2014xwa}. Its main goals include the investigation and mapping of the phase diagram of strongly interacting matter, as well as measuring cross sections of processes relevant for cosmic rays and neutrino physics. In this paper, we are focusing on mapping the QCD phase diagram. In order to accomplish this, NA61/SHINE performs measurements of different collision systems at multiple energies. The experiment provides excellent tracking down to $p_T = 0$ GeV/\textit{c}. This performance is achieved by using four large Time Projection Chambers (TPC's), which cover the full forward hemisphere. The experiment also features a modular calorimeter, called the Projectile Spectator Detector. It is located on the beam axis, after the TPC's, and measures the forward energy which determines the collision centrality of the events. A setup of the NA61/SHINE detector system is shown in Fig.~\ref{f:na61etup}.

The search for the critical endpoint (CEP) and investigation of the QCD phase diagram requires analysis at different temperatures and baryon-chemical potentials. To study, we need to map the phase diagram using different system sizes at various energies. NA61/SHINE investigations cover several beam momenta (13\textit{A}, 20\textit{A}, 30\textit{A}, 40\textit{A}, 75\textit{A} and 150\textit{A} GeV/\textit{c}) and collision systems (p+p,p+Pb,Be+Be,Ar+Sc,Xe+La,Pb+Pb). In this paper, we describe the femtoscopic correlations of identical pions emitted from central Ar+Sc collisions at beam momentum of 150\textit{A} GeV/\textit{c}. This field is often called femtoscopy as it reveals the femtometer scale structure of particle production.

\section{Femtoscopy with L\'evy shaped sources}

The method of quantum-statistical (Bose-Einstein) correlations is based on the work of R. Hanbury Brown and R. Q. Twiss (HBT)~\cite{HanburyBrown:1956bqd}, who applied it first in astrophysical intensity correlation measurements. The method was developed to determine the apparent angular diameter of stellar objects. Shortly afterwards, a similar quantum-statistical method was applied in momentum correlation measurements for proton-antiproton collisions~\cite{Goldhaber:1959mj,Goldhaber:1960sf} by Goldhaber and collaborators. Their objective was to understand pion-pion correlations and gain information on the radius, $R$, of the interaction volume in high-energy particle collisions. The key relationship for measuring Bose-Einstein correlations shows that the spatial momentum correlations, $C(q)$, are related to the properties of the particle emitting source, $S(x)$, that describes the probability density of particle creation for a relative coordinate $x$ as:
\begin{equation}
C_2(q) \cong 1 + | \tilde{S}(q) |^2,
\end{equation}
where $\tilde{S}(q)$ is the Fourier transform of $S(x)$, and $q$ is the relative momentum of the particle pair (with the dependence on the average momentum, $K$, of the pair suppressed and described in more detail in~\cite{NA61SHINE:2023qzr}). The usual assumption for the shape of the source based on the central limit theorem, is a Gaussian. However, such Gaussian shaped sources lead to Gaussian correlation functions. A more general assumption is the L\'evy distribution~\cite{Csorgo:2003uv, Metzler:1999zz}. It exhibits a power-law tail and includes a Gaussian limit, as well. Correlation functions based on this approach have been shown to describe data from different experiments, such as LEP~\cite{L3:2011kzb}, RHIC~\cite{PHENIX:2017ino}, and LHC~\cite{CMS:2017mdg,CMS:2022cvh} quite well. Several phenomena could explain the appearance of L\'evy shaped sources. The non-Gaussianity of the source could be attributed to critical fluctuations and the emergence of spatial correlations on a large scale, which may indicate the existence of similar sources with power-law tails~\cite{Csorgo:2005it}. Further reasons include the fractal structure of QCD jets~\cite{Csorgo:2004sr}. 

In this paper, the measured femtoscopic correlation (including final state interaction) with spherically symmetric L\'evy distributions is defined as:
\begin{equation}
\mathcal{L}(\alpha,R,\vec{r})=\frac{1}{(2\pi)^3} \int d^3\vec{\zeta} e^{i\vec{\zeta} \vec{r}} e^{-\frac{1}{2}|\vec{\zeta} R|^{\alpha}},
\end{equation}
where \textit{R} is the L\'evy scale parameter and $\alpha$ defined as the L\'evy stability index. In addition, $\vec{\zeta}$ is the three-dimensional integration variable with dimensions of MeV/$c$ and $\vec{r}$ is the vector of spatial coordinates. There are two special cases where the distribution can be expressed analytically. One such case is, the already mentioned, Gaussian distribution for $\alpha = 2$. Besides this, the $\alpha=1$ case leads to a Cauchy distribution. An important difference between L\'evy distributions and Gaussians is the presence of a power-law tail $ \sim r^{-(d-2+\alpha)}$ in case of  $\alpha < 2$, where \textit{d} represents the number of spatial dimensions. With the assumption of L\'evy sources, the femtoscopic correlation functions can be expressed in the following way:
\begin{equation}
C_2(q) = 1 + \lambda \cdot e^{-(qR)^\alpha} .\label{e:levyC0}
\end{equation}
$C_2(q)$ is a stretched type of exponential, where the $\lambda$ intercept parameter is defined as:
\begin{equation}
 C_2(q\rightarrow 0) = 1 + \lambda .
\end{equation}
At vanishing relative momentum, the correlation function has a value of $1+\lambda$. This value is not accessible in the measurements and extrapolation from the region, when two tracks are experimentally resolved, is needed. However, it is commonly observed that the intercept parameter $\lambda$ is less than 1. The core-halo model, explained in Refs.~\cite{Csorgo:1999sj,Csorgo:1995bi}, can provide some insights into this parameter.

The model assumes that the source $S$ is made up of two parts, the core and the halo ($S_{\textnormal{core}}$ and $S_{\textnormal{halo}}$), respectively. The core contains pions created directly from hadronic freeze-out or from extremely short lived (strongly decaying) resonances. The halo consists of pions created from longer-lived resonances and the general background. It may extend to thousands of femtometers, while core part has a size of around a few femtometers. In this picture, the $\lambda$ parameter turns out to be connected to the ratio of the core and the halo as:
\begin{equation}\label{e:corr_str}
\lambda = \left(\frac{N_{\rm{core}}}{N_{\rm{core}}+N_{\rm{halo}}}\right)^2.
\end{equation}
Then, one can modify the correlation function to take the effect of the halo into account, by utilizing the Bowler-Sinyukov method~\cite{Sinyukov:1998fc,Bowler:1991vx} as:
\begin{equation}
C_2(q) = 1-\lambda+ \lambda\cdot(1+e^{-|qR|^\alpha}).\label{e:bowlersinyukov}
\end{equation}
The halo part contributes at very small values of relative momenta, \textit{q}. Therefore, it does not affect the source radii of the core part~\cite{Maj:2009ue}. 


It is well known that critical points are characterized by critical exponents. One, in particular, is related to spatial correlations by the exponent denoted as $\eta$. The appearance of the parameter can be explained by the second-order phase transition at the CEP, where fluctuations will appear at all scales causing the spatial correlation function to exhibit a power-law tail with an exponent of $-(d-2+\eta)$, with \textit{d} denoting the dimension. The L\'evy exponent, $\alpha$, is the exponent in the case of, previously defined, L\'evy distributed sources and will also exhibit a power-law tail, with an exponent of $-d+2-\alpha$~\cite{Csorgo:2008ayr}. Hence, $\alpha$ was suggested to be directly related to or being explicitly equal to, the critical exponent, $\eta$~\cite{Csorgo:2005it}, in absence of other phenomena affecting the source shape. This is the basis of the idea connecting $\alpha$ and $\eta$. However, the L\'evy-shape of the source can be attributed to several different factors besides critical phenomena, including QCD jets, anomalous diffusion, critical phenomena, and others~\cite{Metzler:1999zz,Csorgo:2003uv, Csorgo:2004sr, Kincses:2022eqq, Korodi:2022ohn}. Hence, while a non-monotonic behavior of $\alpha$ is expected near the critical point, a detailed understanding of the collision energy and system size dependence of $\alpha$ is needed to draw conclusions about the critical point.

It has been suggested that the universality class of QCD is the same as that of the 3D Ising model~\cite{Halasz:1998qr,Stephanov:1998dy}. The value of $\eta$ in the 3D Ising model is 0{.}03631 $\pm$ 0.00003\cite{El-Showk:2014dwa}. An alternative solution is to use the universality class of the 3D Ising model with a random external field, which yields an $\eta$ value of 0{.}50 $\pm$ 0.05~\cite{Rieger:1995aa}. The statements mentioned suggest that $\alpha$ would decrease to 0{.}50, or below, at the vicinity of the CEP. To confirm this, measurements of $\alpha$ are needed in different collision systems at various energies.

In this analysis, we are dealing with like-charged particles that are influenced by Coulomb repulsion. The final state Coulomb effect has been neglected in the previously defined correlation function. Thus, the correlation function in Eq.~\eqref{e:bowlersinyukov} that lacks this effect will be denoted as $C^0_2(q)$ from now on. The correction necessitated by this effect can be done by simply taking the ratio of $C^{\textnormal{Coul}}_2$ and $C^0_2(q)$: 
\begin{align}\label{e:coulcorr_equation}
K_{\textnormal{Coulomb}}(q) = \frac{C^{\textnormal{ Coul}}_2(q)}{C^0_2(q)} ,
\end{align}
where $C^{\textnormal{Coul}}_2(q)$ is the interference of solutions of the two-particle Schr\"odinger equation; with a Coulomb-potential~\cite{Kincses:2019rug,Csanad:2019lkp}. The numerator in Eq.~\eqref{e:coulcorr_equation} cannot be calculated analytically and requires a large numerical effort to estimate.

An approximate formula for $C^{\textnormal{Coul}}_2(q)$ was obtained in Ref.~\cite{CMS:2017mdg} for the case of Cauchy-shaped sources. However, a more precise treatment is required due to our assumption of L\'evy-shaped sources. We are utilizing a new method in our analysis for estimating the effect of Coulomb repulsion. The treatment includes the numerical calculation presented in Refs.~\cite{Csanad:2019cns, Csanad:2019lkp}, the parametrization of its results, and, finally; the parametrization of the dependence of the physical parameters \textit{R}, $\lambda$, and $\alpha$. Thus Eq.~\eqref{e:bowlersinyukov} is modified as:
\begin{equation}
C_2(q) = N\cdot\left(1-\lambda+\lambda\cdot (1+e^{-|qR|^\alpha})\cdot K_{\textnormal{Coulomb}}(q)\right),\label{e:fittingformula}
\end{equation}
where $N$ is introduced as normalization parameter and  $K_{\textnormal{Coulomb}}(q)$ denotes the Coulomb correction.

It is important to highlight that the Coulomb correction is calculated in the pair-center-of-mass (PCMS) system, while the measurement is often done in the longitudinally co-moving system (LCMS). The assumption of Coulomb correction in the one-dimensional HBT in LCMS picture is that the shape of the source is spherical, i.e. $R_{\textnormal{out}} = R_{\textnormal{side}} = R_{\textnormal{long}} = R \equiv R_{\textnormal{LCMS}}$. The shape of the source, however, is spherical in the LCMS and not in the PCMS. Therefore, an approximate one-dimensional PCMS size parameter is needed. A study was done, where an average PCMS radius of 
\begin{equation}
\overline{R}_{\textnormal{PCMS}} = \sqrt{\frac{1-\frac{2}{3}\beta^2_{\textnormal{T}}}{1-\beta^2_{\textnormal{T}}}} \cdot R
\end{equation}
was calculated~\cite{Kurgyis:2020vbz}, with $\beta_{\textnormal{T}} = \frac{K_{\textnormal{T}}}{m_{\textnormal{T}}}$.

\section{Measurement details}
For the measurements this paper is based on, we analyzed Ar+Sc collisions at 150\textit{A} GeV/\textit{c} beam momentum in the 0-10\% most central events. 
The available data set contains around 2.7 million events, which was reduced to around 700 000 events in the analysis. The following paragraph describes the various event, track and pair selection performed on the data. As mentioned, we have selected 0-10\% of the most central events by measuring the energy contained in the projectile remnants, with the Projectile Spectator Detector (PSD). We have also selected events where no off-time beam particle was detected. These are particles which come within the drift time of the chambers which are not emptied out. Furthermore, all the events chosen have between $\pm 10$ cm as that is the maximal distance between the main vertex \textit{z} position and the center of the scandium target. Particle identification was handled by using d$E$/d$x$ energy deposit in the TPC gas. The tracks were extrapolated to the interaction plane and matched with the distance against the interaction point. If the distance was $\leq$ 4 cm in the horizontal plane and $\leq$ 2 cm in the vertical plane, the track was kept. Moreover, track splitting was handled by selecting the ratio of the total number of reconstructed points on the track to the potential number of points to be between 0{.}5 and 1{.}0. Finally, to counter track merging, we have used a selection on the momentum space distance between the two tracks. It uses non-standard momentum coordinates, $s_x = p_x/p_{xz}$, $s_y = p_y/p_{xz}$ and $\rho = 1/p_{xz}$. 

\begin{figure}
\centering
\includegraphics[width=1.0\textwidth]{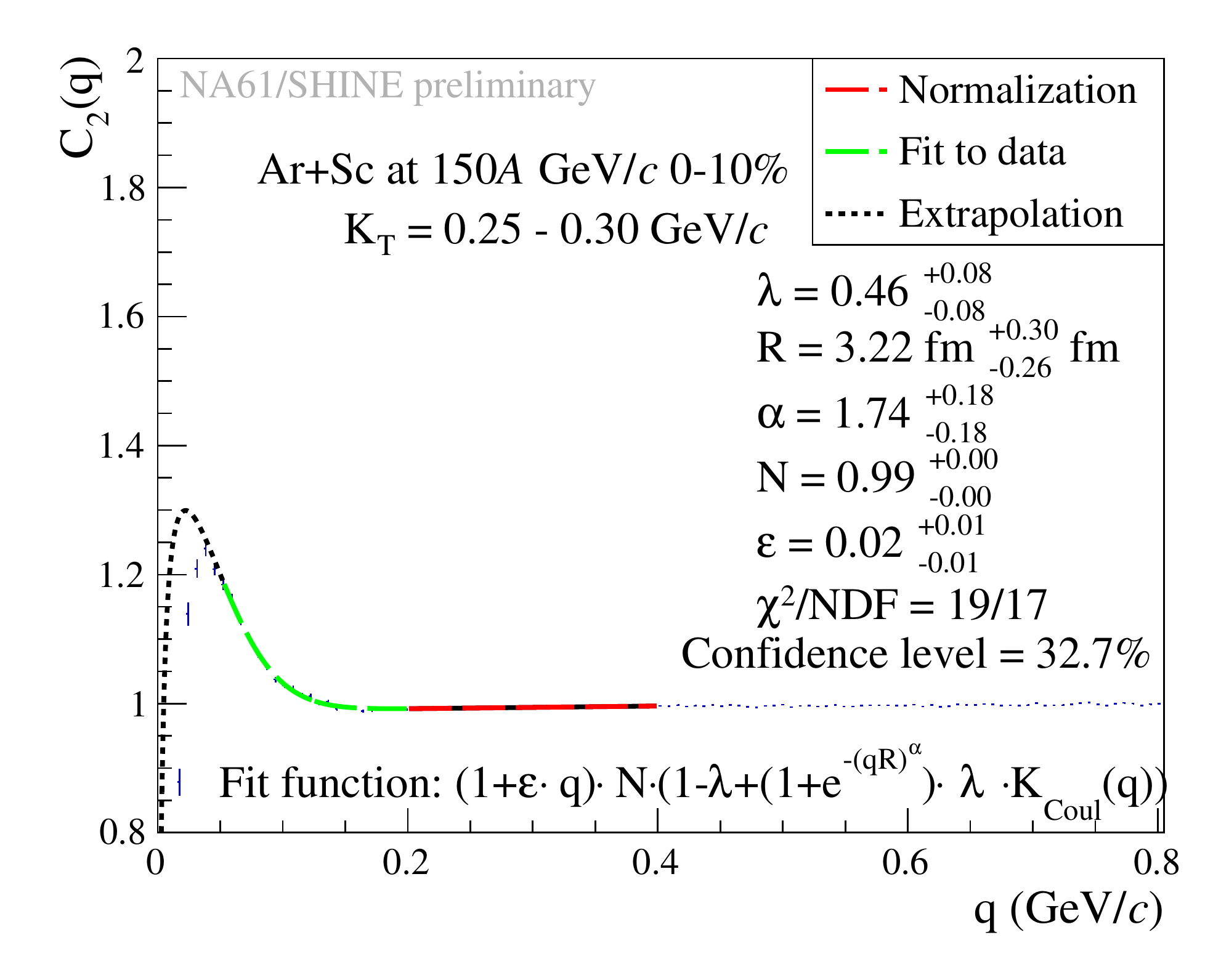}
\caption{Example fit with Bose--Einstein correlation function at $K_{\textnormal{T}} = 0.25$--$0.30$~GeV/\textit{c} for the sum $\left(\pi^++\pi^+\right) + \left(\pi^-+\pi^-\right)$. Blue points with error bars represent the data, the green dash-dotted line shows the fitted function with Coulomb correction given by Eq.~\eqref{e:fitfunc} within the range of 0.0525~GeV/\textit{c} to 0.2~GeV/\textit{c} and red dashed line represent normalization to background 0.2~GeV/\textit{c} to 0.4~GeV/\textit{c}. In the low-$q$ region, the black dotted line indicates the extrapolated function outside of the fit range due to prominent detector resolution effects, mentioned in the text.}
\label{fig:examplefit}
\end{figure}

We then analyzed the combination of negative pion pairs and positive pion pairs. It is important to note that this analysis was done with a one-dimensional relative momentum variable $q$, calculated in LCMS. These pairs were sorted into eight $K_{\textnormal{T}}$ (average transverse momentum  of the pair) bins in the range of 0-450 MeV/\textit{c}. In each momentum bin, the relative momentum distribution of coincident pion pairs were obtained. Let us call this the actual pair momentum difference distribution, $A(q)$. $A(q)$ contains quantum-statistical correlations, as well as many other residual effects related to kinematics and acceptance. The effects can be removed by constructing a combinatorial background pair distribution, denoted as $B(q)$, which is measured in the same $K_{\textnormal{T}}$ or $m_{\textnormal{T}}$ intervals as the $A(q)$ distribution. The method we use involves randomly selecting the same number of particles as the multiplicity of the actual event. The selected particles are from other events of similar parameters and each is selected from a different event. Let us call the pair momentum difference distribution made from this method as, the background distribution, $B(q)$. This, by construction, enables us to have an uncorrelated pool of events. Then the correlation function is calculated as
\begin{equation}
C_2(q) = \frac{A(q)}{B(q)}\cdot \frac{\int_{q_1}^{q_2}B(q)dq}{\int_{q_1}^{q_2}A(q)dq}\;,\label{e:corrfuncAB}
\end{equation}
in a $[q_1,q_2]$ range where quantum-statistical correlations are not expected. Upon adding the contribution from the background, the previously defined Eq.~\eqref{e:fittingformula} is modified as follows:
\begin{equation}\label{e:fitfunc}
C(q) =  N\cdot(1 + \varepsilon\cdot q)\cdot\left(1-\lambda + \lambda\cdot\left(1+e^{-(qR)^{\alpha}}\right)\cdot K_{\textrm{Coulomb}}(q)\right),
\end{equation}
with $N$ being a normalization parameter responsible for the proper normalization of the $A(q)/B(q)$ ratio, $\varepsilon$ describing the linearity of background, and $K_{\textrm{Coulomb}}(q)$ being the Coulomb correction. We then use this formula to our measured $C_2(q)$ as shown in Fig.~\ref{fig:examplefit}. We then use this formula to our measured $C_2(q)$ as shown in Fig.~\ref{fig:examplefit}. To determine the fitting range where the effects of detector resolution do not play a significant role, we have used EPOS simulation~\cite{Pierog:2009zt} with GEANT3 for particle propagation~\cite{Geant3}. Note that in the low-$q$ region, the fit does not describe the data. This can be explained, according to Monte Carlo simulations of the detector response, by the limited resolution of pairs with small relative momentum.

\section{Results}
The three physical parameters $\left( \alpha, R \textnormal{, and } \lambda \right)$ were measured in eight bins of pair transverse momentum, $K_{\textnormal{T}}$. The three mentioned parameters were obtained through fitting the measured correlation functions with the formula shown in Eq. \eqref{e:fitfunc}. The results were investigated regarding their transverse momentum dependence. In the following, we report on the transverse mass dependence of $\alpha$, $\lambda$, and $R$; where transverse mass is expressed as $m_{\textnormal{T}}=\sqrt{m_{\pi}^2c^4+K_{\textnormal{T}}^2c^2}$, with $m_{\pi}$ being the pion mass.

\begin{figure}[H]
\centering
\includegraphics[width=1\textwidth]{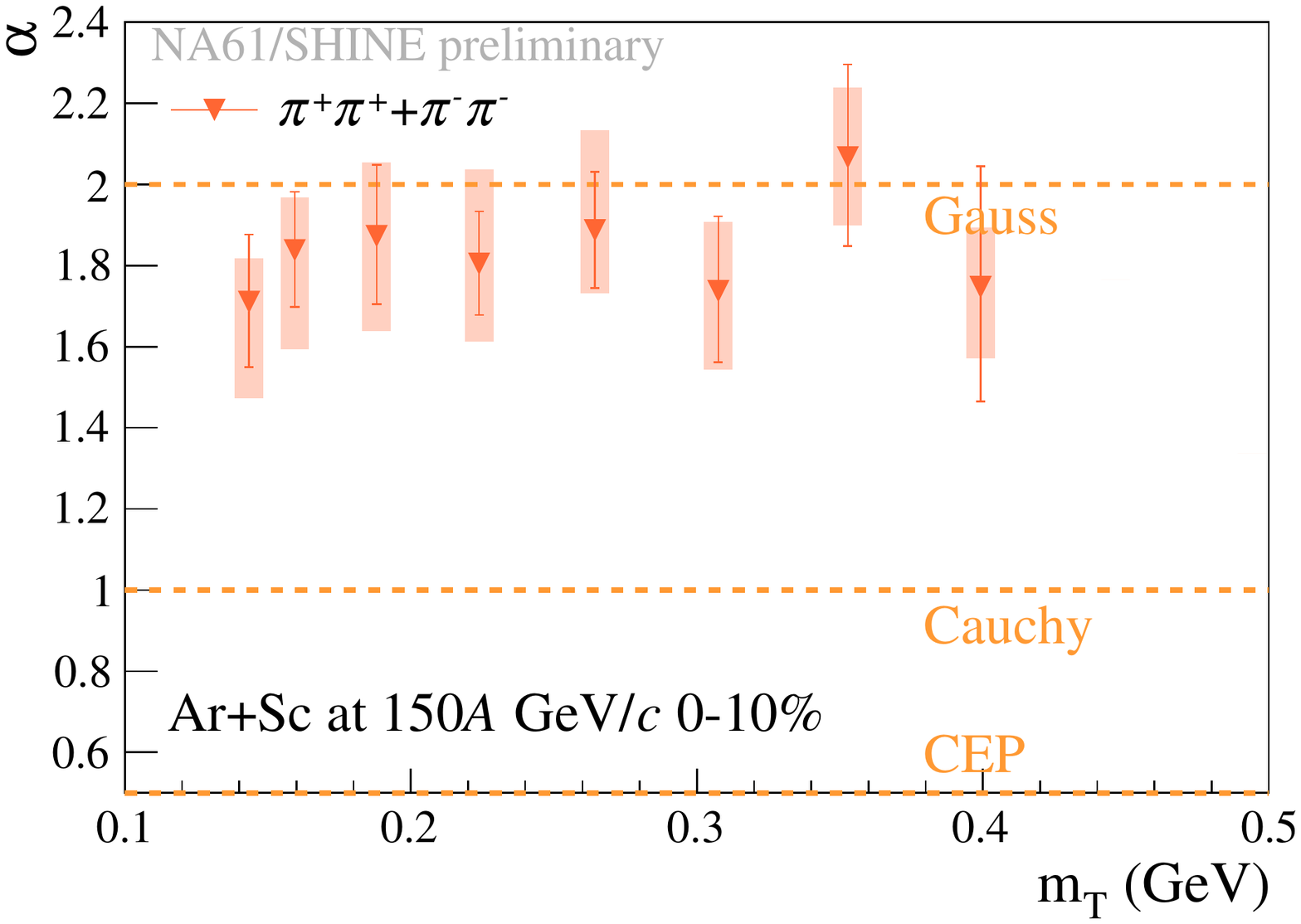}
\caption{The L\'evy stability index, $\alpha$, for $0$--$10$\% central Ar+Sc at 150\textit{A} GeV/\textit{c}, as a function of $m_{\textnormal{T}}$.
      Special cases corresponding to a Gaussian ($\alpha=2$) or a Cauchy ($\alpha=1$) source are shown, as well as $\alpha=0.5$, the conjectured value corresponding to the critical endpoint. Boxes denote systematic uncertainties, bars represent statistical uncertainties.}
\label{fig:alpha}
\end{figure}

As explained above, the shape of the source is often assumed to be Gaussian. The L\'evy stability exponent, $\alpha$, can be used to extract the shape of the tail of the source. Our results, shown in Fig.~\ref{fig:alpha}, yield values for $\alpha$ between 1.5 and 2.0, which imply a source closer to the Gaussian shape than the one in Be+Be collisions~\cite{NA61SHINE:2023qzr}, but are still significantly lower than the $\alpha = 2$ (Gaussian) case. The observed $\alpha$ parameter is also significantly higher than the conjectured value at the critical point ($\alpha = 0.5$). Altogether, these results suggest that measured correlation functions align with the assumption of a L\'evy source, indicating that it is more advantageous over the Gaussian assumption. Further studies are ongoing at NA61/SHINE, using different collision energies and system sizes, in order to map the evolution of the L\'evy stability index, $\alpha$, as a function of collision energy and system size.

\begin{figure}[H]
\centering
\includegraphics[width=1\textwidth]{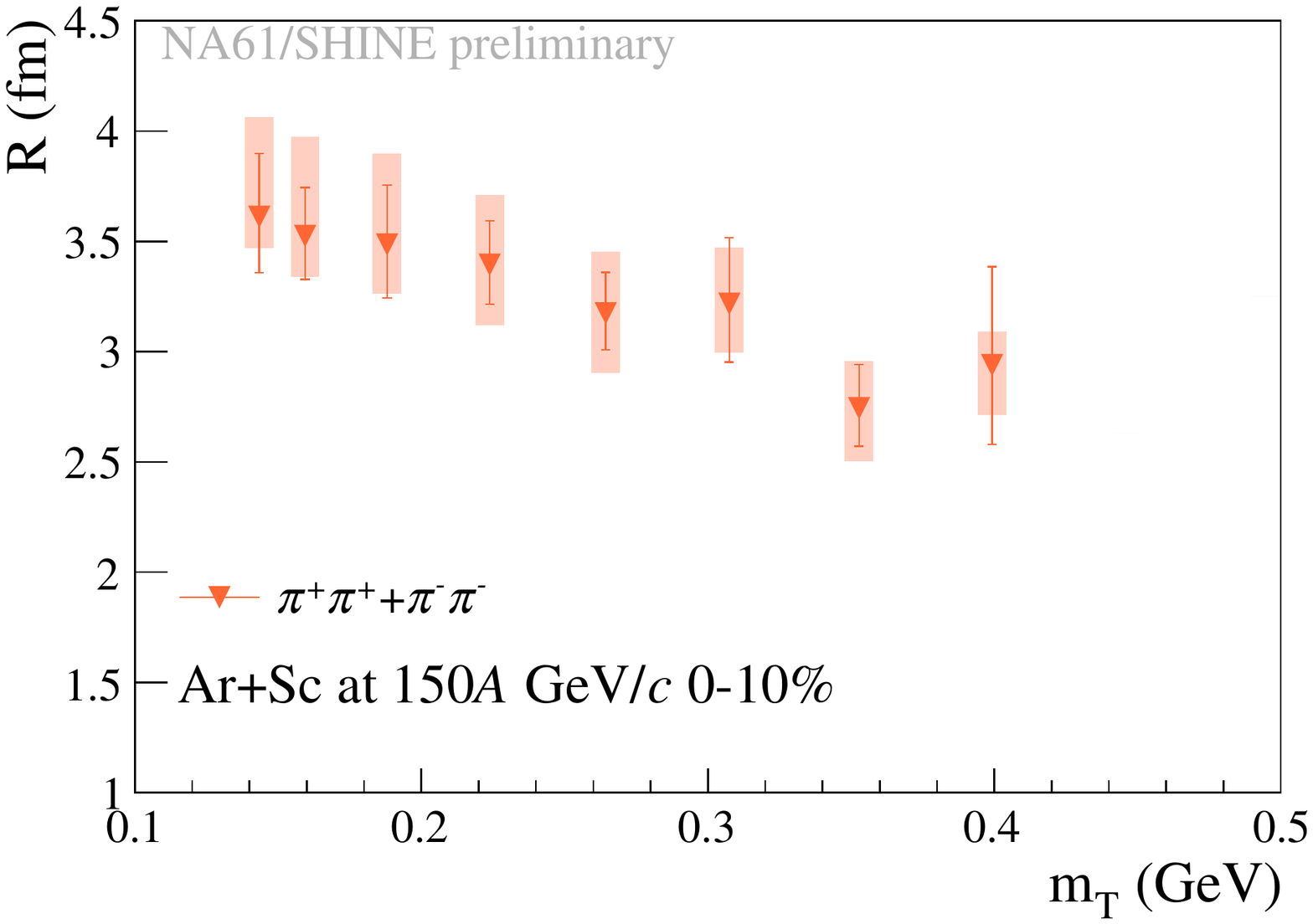}
\caption{The radial scale parameter, $R$, for 0--10\% central Ar+Sc at 150\textit{A} GeV/\textit{c}, as a function of $m_{\textnormal{T}}$. Boxes denote systematic uncertainties, bars represent statistical uncertainties.}
\label{fig:R}
\end{figure}

As a second parameter, let us look at the L\'evy scale parameter, $R$, visible in Fig.~\ref{fig:R}. It determines the length of homogeneity of the pion emitting source. The parameter $R$ depends on the transverse-mass as $1/\sqrt{m_{\textnormal{T}}}$. This can be derived using simple, hydrodynamical predictions for Gaussian sources~\cite{Csorgo:1995bi,Csanad:2009wc}:
\begin{equation}
    \frac{1}{R^2_{\textnormal{HBT}}} = \frac{1}{R^2_{\textnormal{geom}}} + u^2_{\textnormal{T}}\cdot\frac{m_{\textnormal{T}}}{T_0},
\end{equation}
where $u^2_T$ is the average, transverse expansion and $T_0$ is the hadronisation temperature. In our case, rather surprisingly, despite the non-Gaussian nature of the source, this formula works in describing the measured femtoscopic radii. More precisely, as mentioned above, observing an $R \sim 1/\sqrt{m_{\textnormal{T}}}$ is particularly interesting as this type of $m_{\textnormal{T}}$ dependence should rise in case of Gaussian sources ($\alpha=2$)\cite{Sinyukov:1994vg}. It is not entirely clear why this happens, the indicated $m_{\textnormal{T}}$ dependence could form in the QGP or at a later stage. This phenomenon was also observed at RHIC~\cite{PHENIX:2017ino} and in simulations at RHIC and LHC energies~\cite{Kincses:2022eqq, Korodi:2022ohn}.

\begin{figure}[H]
\centering
\includegraphics[width=1\textwidth]{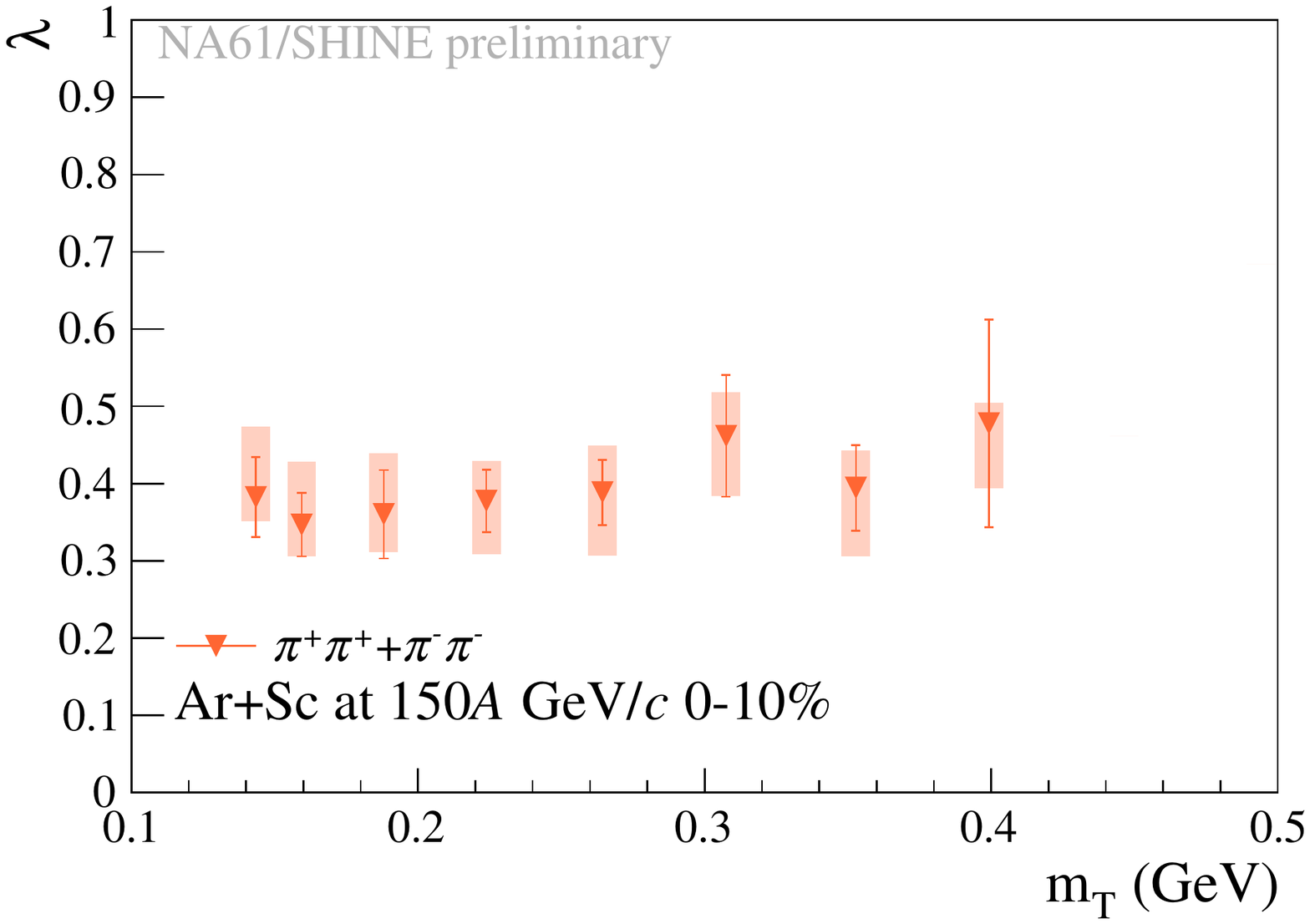}
\caption{The correlation strength parameter, $\lambda$, for 0--10\% central Ar+Sc at 150\textit{A} GeV/\textit{c}, as a function of $m_{\textnormal{T}}$. Boxes denote systematic uncertainties, bars represent statistical uncertainties.}\label{fig:lambda}
\end{figure}

The final parameter being investigated is the intercept (also known as the correlation strength) parameter, $\lambda$, defined in Eq.~\eqref{e:corr_str}. The dependence of $\lambda$ on $m_{\textnormal{T}}$ is shown in Fig.~\ref{fig:lambda}. One may observe a slight dependence on $m_{\textnormal{T}}$, but this can still can be considered constant in the investigated range. When compared to measurements from RHIC Au+Au collisions~\cite{PHENIX:2017ino,Vertesi:2009wf, STAR:2009fks} and from SPS Pb+Pb interactions~\cite{Beker:1994qv,NA49:2007fqa}, an interesting phenomenon is observed. At the energies of SPS, there is no visible ``hole'' at lower $m_{\textnormal{T}}$ values, but at RHIC energies, the ``hole'' appears at $m_{\textnormal{T}}$ values of around a few hundred MeV.  This ``hole'' was interpreted in Refs.~\cite{PHENIX:2017ino} and~\cite{Vance:1998wd} to be a sign of in-medium mass modification. The results presented in Fig.~\ref{fig:lambda}, at the given statistical precision, do not indicate the presence of such a low-$m_{\textnormal{T}}$ hole. This trend might imply that this phenomenon can be turned off at SPS. Furthermore, it can be highlighted that the $\lambda$ values we obtained are significantly below unity. A possible answer can be given by the halo part of the core-halo model. It may indicate that a significant fraction of pions are the decay products of long-lived resonances.

\section{Conclusions}
In the report above, we discussed the NA61/SHINE measurement of one-dimensional, identified, two-pion, femtoscopic correlation functions; in the 0-10\% most central Ar+Sc collisions at 150\textit{A} GeV/\textit{c}. We discussed the transverse mass dependencies of the L\'evy source parameters. Results on the L\'evy scale parameter, $\alpha$, showed a significant deviation from Gaussian sources and are not in the vicinity of the conjectured value at the critical point. The L\'evy scale parameter, $R$, shows a visible decrease with $m_{\textnormal{T}}$. The correlation strength parameter, $\lambda$, does not show any significant $m_{\textnormal{T}}$ dependence, but, maps different patterns at RHIC and similar trends at SPS energies. With these results at hand, we plan to measure Bose-Einstein correlations in larger systems, as well as at smaller energies, to continue mapping the phase diagram of the strongly interacting matter.

\vspace{6pt} 




\funding{This research was supported by the \'UNKP-22-3 New National Excellence Program of the Ministry for Culture and Innovation from the source of the National Research, Development and Innovation Fund, and the NKFIH OTKA K-138136 grant.}

\dataavailability{The data presented in this study are available on request from the corresponding author. The data are not publicly available.} 

\acknowledgments{The author would like to thank the NA61/SHINE collaboration.}

\conflictsofinterest{The author declares no conflict of interest.}

\abbreviations{Abbreviations}{
The following abbreviations are used in this manuscript:\\

\noindent 
\begin{tabular}{@{}ll}
QCD & quantum chromodynamics\\
CERN & Conseil européen pour la recherche nucléaire\\
SPS & Super Proton Synchrotron\\
HBT & Hanbury Brown and Twiss\\
BE & Bose-Einstein\\
CEP & critical endpoint\\
NA61/SHINE & North Area 61 / SPS Heavy Ion and Neutrino Experiment\\
LCMS & Longitudinally Co-Moving System
\end{tabular}
}

\begin{adjustwidth}{-\extralength}{0cm}

\reftitle{References}


\bibliography{HBT_NA61_Zimanyi}

\PublishersNote{}
\end{adjustwidth}
\end{document}